\documentstyle[preprint,prb,aps,epsf]{revtex}
\begin{document}
\tightenlines
\draft

\title{Phase separation and the segregation principle in the infinite-$U$
spinless Falicov-Kimball model}
\author{J. K. Freericks$^{\dagger}$, Ch. Gruber$^*$, and N. Macris$^*$}
\address{$^{\dagger}$Department of Physics, Georgetown University, 
Washington, DC 20057 USA\\
$^*$Institut de Physique Th\'eorique, Ecole Polytechnique F\'ed\'erale de 
Lausanne,\\
PHB-Ecublens, CH-1015, Lausanne, Switzerland}
\maketitle
\begin{abstract}
The simplest statistical-mechanical
model of crystalline formation (or alloy formation)
that includes electronic degrees of freedom is solved exactly 
in the limit of large spatial dimensions and infinite interaction
strength.  The solutions contain both
second-order phase transitions and first-order phase transitions (that
involve phase-separation or segregation) which are likely to illustrate
the basic physics behind the static charge-stripe ordering in cuprate systems.  
In addition, we find the spinodal-decomposition temperature satisfies an
approximate scaling law.
\end{abstract}
\widetext
\section{Introduction}

The most fundamental problem in solid-state physics is to understand
why elements (and most compounds) crystallize in ordered periodic 
structures, for this forms the basis of
all of solid-state physics.  While it is well known that
the driving principle behind this ordering is a lowering of the
ground-state energy of the material, and there has been significant progress
with {\it ab initio} methods to predict the ground-state properties of these
ordered phases in real materials, there still are no exactly solvable
models for crystal formation that describe the statistical-mechanical
mechanism behind the ordering of the electrons and ions on a periodic
lattice.  Furthermore, it is not understood what the physical mechanisms are
that are necessary for creating a crystallized state. This crystallization
problem is ubiquitous; it also describes the statistical mechanics behind
binary-alloy formation or phase separation since the two problems can be
mapped onto each other (as described below), and it 
may also describe the physics behind charge-stripe formation in the cuprates.

It may sound surprising that no solvable statistical-mechanical model for
crystallization exists, since a statistical-mechanical model for
magnetic order has been known ever since Onsager solved
the two-dimensional Ising model\cite{onsager}.  Onsager's solution
produced a paradigm for understanding phase
transitions in many different physical systems and provided a textbook
example of much of the theory behind modern critical phenomena.  In fact,
Lee and Yang\cite{lee_yang} modified the Ising model to consider the 
magnetic order in an external magnetic field, and mapped the problem onto a
lattice gas, where the up spins denoted sites occupied by ions, and the down
spins denoted empty sites.  Onsager's method of solution does not extend to
the case of a finite magnetic field, so no exact results are known for
the lattice gas, except in the case where the number of ions equals
one half the number of lattice sites, which corresponds to the zero-field
case.  These models of crystallization neglect the electronic degrees of
freedom of the valence electrons, and hence are not directly applicable
to real materials such as metals and alloys.  

It turns out
that the Ising model, and many other models for magnetism, simplify when
they are examined in high dimensions.  In fact, the Ising model is solved by
a static mean field theory in four and higher dimensions.  
A similar situation is expected for electronic problems, except they remain
nontrivial even in the infinite-dimensional limit\cite{metzner_vollhardt,GKKR}.
Metzner and Vollhardt showed that the electronic problem
requires a dynamical mean-field theory for its solution
in infinite dimensions.  Furthermore,
a wide range of evidence indicates that this dynamical mean-field theory
provides a quantitative approximation to the solutions of correlated electron
problems in three dimensions (at least if one is not too close to a critical
point).  In fact, it is precisely the nonuniversal properties (such as a
transition temperature) that the dynamical mean-field theory 
determines accurately, and it's solution provides a wealth of
information on the qualitative behavior of the model studied.
We employ the dynamical mean-field theory here to produce an
exact solution of the crystallization problem which includes the electronic
degrees of freedom.  

The simplest model that can describe crystallization and include electronic
degrees of freedom is the spinless 
Falicov-Kimball model\cite{falicov_kimball} which consists of two kinds
of particles: localized ions and itinerant (spinless) electrons.
The localized ions ($w_i=0$ or 1)
occupy sites on a lattice in real space with an energy $E$, and the electrons 
can hop (with a
hopping integral $-t^*/[2\sqrt{d}]$) between
neighboring lattice sites.  In addition, there is a screened Coulomb
interaction $U$ between electrons and ions that occupy the same lattice site.
Since the electrons do not interact with each other, the ``spin'' degree of
freedom is unimportant, and is neglected.  The Hamiltonian is
\begin{equation}
H=-\frac{t^*}{2\sqrt{d}}\sum_{<i,j>}c_i^{\dagger}c_j+E\sum_iw_i
+U\sum_ic_i^{\dagger}c_iw_i,
\end{equation}
with $c_i^{\dagger}$ $(c_i)$ the creation (annihilation) operator for
electrons at site $i$, and $w_i$ denoting the ion occupancy at site $i$.
We use $t^*=1$ as the energy scale.

The Falicov-Kimball model can be viewed as a simplified approximation to
a real material in a variety of ways.  If the material has a single valence
electron, and only one electronic band lies near the Fermi level, then
the crystallization problem would correspond to the case where
the electron and ion concentrations ($\rho_e$ and $\rho_i$) are the same
(which is called the neutral case), since one electron is donated
by each ion.   
If, instead, there are many bands near the Fermi level, then one can map
the combined bands into a single ``effective'' band which will have 
an electron filling determined by the average filling of the electrons in
the most important band.  In this case, each ion may donate only a fraction of 
an electron to the crystal, because the rest of the electron goes into other
hybridized bands that lie close to the Fermi level.  Hence, one may find
it useful to also consider nonneutral cases for the crystallization 
problem, where the electron and ion concentrations are not equal.
This model can also be mapped
onto the binary  alloy problem, where a site occupied by an ion is mapped to
a site occupied by an $A$ ion and a site unoccupied by an ion is mapped
to a site occupied by a $B$ ion, and the screened Coulomb interaction is
mapped to the difference in the site energies for electrons on an $A$
ion versus on a $B$ ion.  

As it stands, the Falicov-Kimball model doesn't appear to be a many-body
problem at all, since the ions are localized and do not move, which implies that
the quantum-mechanical problem for the electrons
can be solved by diagonalizing a single-particle
problem of an electron moving in the potential determined by the 
given configuration of the ions $\{w_i\}$.  The many-body problem aspects 
enter by taking an annealed average over all possible ion configurations with 
the chosen ion concentration.  This produces long-range
interactions between the ions, that can cause them to order or phase
separate at low temperatures.

Much is already known about the physics of the Falicov-Kimball model (as 
reviewed by Gruber and Macris\cite{GM}).  In
the neutral case where each particle concentration equals 1/2,  
Lieb and Kennedy \cite{lieb_kennedy} and Brandt and Schmidt 
\cite{brandt_schmidt}  proved that the system always orders
in an alternating ``chessboard'' phase at a finite transition temperature
in all dimensions greater than 1.  This ordered phase can
be interpreted as the transition from a high-temperature homogeneous
(liquid/gas) phase to a low-temperature ordered (solid) phase.
The appearance of a low-temperature ordered phase follows as a consequence of
the Pauli exclusion principle, since Lieb and Kennedy also showed that if
the itinerant particles were Bosons instead of electrons, they would clump
together and not form a periodically ordered ground state.   

The Falicov-Kimball model is expected to be in the same universality class as 
the Ising model, but, because of the electronic degrees of freedom, one
needs to solve the full statistical model to determine the ``effective
magnetic exchange parameters'' between different lattice sites.
The parameters can be extracted in a systematic expansion if the electronic
kinetic energy (the hopping term) is taken as a 
perturbation,\cite{BKU,DFF,GMMU} but such an
analysis is only valid in the strong-coupling regime, and rapidly becomes
problematic.  {\it It is precisely this complication that has frustrated 
attempts at finding an exact solution to the crystallization problem
when electronic degrees of freedom are introduced.}

The one-dimensional limit of the Falicov-Kimball model has also been
extensively studied.  Here there are no finite-temperature phase
transitions, but the system can have phase transitions in the ground state.
The first attempt at studying the one-dimensional Falicov-Kimball model
proceeded along the lines of {\it ab initio} band-structure calculations
for real materials---a small number of candidate ion configurations
were chosen for the ground state, and a restricted phase diagram was determined
for all structures within the subset \cite{freericks_falicov}.
The numerical solutions produced two conjectures: the first was a result for
the case where $\rho_e\ne 1-\rho_i$, which stated that if
the screened Coulomb interaction $U$ was large enough, then the 
system would segregate into an empty lattice (with no ions and all the
electrons), and
a full lattice (with all the ions and no
electrons).  The second was a generalization of the Peierls instability,
which says that in the small $U$ limit the system will order in such a fashion
that the ions produce a band structure that has a maximal gap at the Fermi
level.  This first conjecture (the segregation principle)
was later proven to be true by Lemberger
\cite{lemberger} while the second conjecture was shown to be false if
the electron concentration was sufficiently far from half-filling.  In
that case, the system would phase separate between the empty lattice,
and an optimally chosen ion structure that had the Fermi level lying
in the gap \cite{freericks_gruber_macris}.

The other limit that has been extensively studied is the large-dimensional
limit where Brandt and Mielsch \cite{brandt_mielsch} provided the
solution of the transition temperature as a function of $U$ for the
half-filled symmetric case.  Their solution involves solving a coupled
set of transcendental equations which display first and second-order
phase transitions.
Freericks \cite{freericks} later showed that the model (on a hypercubic lattice)
also displayed incommensurate order and segregation.

There are two kinds of lattices that are usually investigated in the large
coordination-number limit:  the hypercubic lattice, which is the generalization
of the cubic lattice to large dimensions; and the Bethe lattice, which is
a thermodynamic limit of the Cayley tree when the number of
nearest neighbors becomes large.  The noninteracting band structure for the
hypercubic lattice produces a density of states
that is a Gaussian [$\rho_H(\epsilon)=\exp(-\epsilon^2)/\sqrt{\pi}]$, 
while on the Bethe lattice the density of states is Wigner's semicircle 
$[\rho_B(\epsilon)=\sqrt{4-\epsilon^2}/(2\pi)]$.  The hypercubic density of
states has an infinite bandwidth, but most of the weight lies within a range of 
$\pm2$ about the origin. The Bethe lattice density of states has the same
behavior as a three-dimensional system at the band edge (square-root behavior)
but has no van Hove singularities in the interior of the band.
Because both density of
states are nontrivial, the many-body problem maintains much of it's
rich behavior that arises from the competition between kinetic-energy
effects and interaction-energy effects.  In particular, the Falicov-Kimball
model continues to have phase transitions
in the large coordination number limit, but the transitions have
mean-field theory exponents.

In this contribution, we examine
what happens in the case when the Coulomb interaction 
becomes infinite $U\rightarrow\infty$ (the attractive case is
equivalent to this case through a particle-hole transformation of the electrons,
which carries $\rho_e\rightarrow 1-\rho_e)$.  In this case, the electrons
avoid the sites of the lattice occupied by the ions, so the electron 
concentration varies from zero up to $1-\rho_i$.  We investigated the 
non-unit-density cases, 
where the electron concentration was restricted to $0\le \rho_e <1-\rho_i$.  
In Section II the formalism and results for calculations on the Bethe lattice
are presented.  In Section III, results for the hypercubic lattice are given
and in Section IV we present our conclusions.

\section{Formalism and Results for the Bethe Lattice}

In the thermodynamic limit, the local lattice Green's function is defined to be
\begin{equation}
G_n=G(i\omega_n)=-\int_0^{\beta}d\tau e^{i\omega_n\tau}\frac{Tr<e^{-\beta 
(H-\mu N)}
T_{\tau} c(\tau)c^{\dagger}(0)>}{Tr<e^{-\beta (H-\mu N)}>},
\label{eq: greendef}
\end{equation}
where $i\omega_n=i\pi T(2n+1)$ is the Fermionic Matsubara frequency,
$\beta=1/T$ is the inverse temperature, $\mu$ is the electron chemical
potential, and $T_{\tau}$ denotes $\tau$-ordering.  The angle brackets in
Eq.~(\ref{eq: greendef}) denote the sum over ionic configurations.
The local Green's function is determined by mapping onto an atomic problem
in a time-dependent field, with the following action
\begin{equation}
S_{at}=\int_0^{\beta}d\tau\int_0^{\beta}d\tau^{\prime}c^{\dagger}(\tau)
G_0^{-1}(\tau-\tau^{\prime})c(\tau^{\prime})+U\int_0^{\beta}d\tau
c^{\dagger}(\tau)c(\tau)w+Ew,
\label{eq: action}
\end{equation}
where $w=0, 1$ is the ion number for the atomic site
and $G_0^{-1}$ is the mean-field or effective-medium
Green's function, which is determined self-consistently (as described below).
The atomic Green's function, with the action in Eq.~(\ref{eq: action}),
is computed to be
\begin{equation}
G_n=\frac{1-\rho_i}{G_0^{-1}(i\omega_n)}+\frac{\rho_i}{G_0^{-1}(i\omega_n)-U},
\label{eq: greenatomic}
\end{equation}
with $\rho_i$ the average ion density $<w>$.  On the other hand, the local 
lattice Green's function satisfies
\begin{equation}
G_n=\int_{-\infty}^{\infty}d\epsilon \frac{\rho(\epsilon)}{i\omega_n+\mu-
\Sigma_n-\epsilon},
\label{eq: greenlocal}
\end{equation}
where $\rho(\epsilon)$ is the noninteracting density of states for the
infinite lattice and $\Sigma_n$ is the self-energy.  The self-consistency 
relation is that the self-energy $\Sigma_n$ in Eq.~(\ref{eq: greenlocal})
must coincide with the self-energy of the atomic problem, i.~e.
\begin{equation}
\Sigma(i\omega_n)=G_0^{-1}(i\omega_n)-G_n^{-1}.
\label{eq: sigmadyson}
\end{equation}
Equations (\ref{eq: greenatomic}), (\ref{eq: greenlocal}), and
(\ref{eq: sigmadyson}) constitute the mean-field theory for homogeneous
phases.  In the limit $d\rightarrow\infty$ Eq.~(\ref{eq: sigmadyson})
is an exact equation for the lattice problem.  We note that for periodic
phases, if they exist, one needs to replace the atomic problem by a more
complicated many-site problem\cite{bethe_periodic}.

These equations are complicated to solve analytically but a simplification 
occurs for $0\le\rho_e\le1-\rho_i$ in the limit $U\rightarrow\infty$.
Indeed, when $U$ is large the spectrum of the Hamiltonian consists of two
bands separated by a gap of order $U$ for electron fillings that satisfy
$0\le\rho_e\le1-\rho_i$.  In this case the chemical potential lies within the
lower band, so that $\mu$ is $O(1)$.  We note that $G_0$ is a function of
$\mu$, and therefore for any finite $\mu$, $1/[G_0(i\omega_n)U]\rightarrow 0$
as $U\rightarrow\infty$.  Then Eq.~(\ref{eq: greenatomic}) becomes
\begin{equation}
G_n=(1-\rho_i)G_0(i\omega_n),
\label{eq: greeninfinity}
\end{equation}
and substituting 
Eq.~(\ref{eq: greeninfinity}) into Eq.~(\ref{eq: sigmadyson}) and solving for 
the self energy, then yields 
\begin{equation}
\Sigma_n=-\frac{\rho_i}{G_n},
\label{eq: sigmainfinity}
\end{equation}
for the relation between the local self energy and the Green's function.
Hence, in the limits $U\rightarrow\infty$ and $d\rightarrow\infty$ the equations
for the homogeneous phase reduce to Eqs.~(\ref{eq: greenlocal}) and
(\ref{eq: sigmainfinity}).

In the case of the
Bethe lattice, $\rho_B(\epsilon)=\sqrt{4-\epsilon^2}/(2\pi)$, for
$-2<\epsilon <2$, so that 
the integral in Eq.~(\ref{eq: greenlocal}) can be performed analytically
\begin{equation}
G_n=\frac{i\omega_n+\mu-\Sigma_n}{2}-\frac{1}{2}
\sqrt{(i\omega_n+\mu-\Sigma_n)^2-4}.
\label{eq: greenbethe}
\end{equation}
Substituting the result from Eq.~(\ref{eq: sigmainfinity}) into 
Eq.~(\ref{eq: greenbethe}) and solving for $G_n$ yields the exact result for the
interacting Green's function in the strongly correlated limit
\begin{equation}
G_n=\frac{i\omega_n+\mu}{2}-\frac{1}{2}\sqrt{(i\omega_n+\mu)^2-4(1-\rho_i)},
\label{eq: greeninteracting}
\end{equation}
where the phase of the square root is chosen so that the Green's function
has the correct sign to it's imaginary part.
This form is identical to that of a noninteracting Green's function 
[Eq.~(\ref{eq: greenbethe}) with $\Sigma_n=0$], with a bandwidth narrowed
from 4 to $4\sqrt{1-\rho_i}$, and containing a spectral weight of $1-\rho_i$
(since the remaining spectral weight is shifted to infinite energies).
This is easiest seen from the interacting density of states, which 
satisfies\cite{vandongen}
\begin{equation}
\rho_B^{int}(\epsilon)=\frac{1}{2\pi}\sqrt{4(1-\rho_i)-\epsilon^2}.
\label{eq: dosbethe}
\end{equation}
Note that in the infinite-interaction-strength limit, we have an analytic
form for the Green's functions, and do not need to iteratively solve 
transcendental equations as is normally done in the finite-$U$ 
case.\cite{brandt_mielsch}
Furthermore, even though the Green's function has the same form as a 
noninteracting Green's function, the self-energy is nontrivial and does
not correspond to a Fermi liquid!

This form for the Green's function fits a rather simple physical picture.
The electron avoids sites occupied by an ion when $U\rightarrow\infty$,
so the number of available sites is reduced by the fraction $1-\rho_i$.
This means, on average, the number of nearest neighbors is reduced by
the same factor, which reduces the bandwidth by $\sqrt{1-\rho_i}$.  The
total spectral weight is also reduced from 1 to $1-\rho_i$, because the
upper band (with $\rho_i$ states) is located at infinite energy.  What is
surprising is that this ``hand-waving'' argument is exact for the Bethe
lattice (we will see below it is a good approximation for the hypercubic
lattice, but is not exact).

The interacting density of states is temperature-independent\cite{temp_ind}
in the local approximation, which means that we can examine the ground state
at $T=0$ to see if the system phase separates, or if the homogeneous
phase is lowest in energy.  The ground state energy for an ion concentration
$\rho_i$ and an electron concentration $\rho_e$ is
\begin{equation}
E(\rho_e,\rho_i)=\int_{-\infty}^{\mu}d\epsilon\rho_B^{int}(\epsilon)\epsilon ,
\label{eq: gs_energy_def}
\end{equation}
with $\mu$ the chemical potential defined by
\begin{equation}
\rho_e=\int_{-\infty}^{\mu}d\epsilon\rho_B^{int}(\epsilon) ,
\label{eq: gs_mu_def}
\end{equation}
and $\rho_B^{int}$ the interacting density of states.  Substituting in the
exact result from Eq.~(\ref{eq: dosbethe}) yields
\begin{equation}
E(\rho_e,\rho_i)=-\frac{4}{3\pi}(1-\rho_i)^{3/2}\left [ 1-
\frac{\mu^2}{4(1-\rho_i)}\right ]^{3/2},
\label{eq: gs_energy}
\end{equation}
and
\begin{equation}
\rho_e=\frac{1-\rho_i}{\pi}\left [ \cos^{-1}
\left ( \frac{-\mu}{2\sqrt{1-\rho_i}}\right )
+\frac{\mu}{2\sqrt{1-\rho_i}}\sqrt{1-\frac{\mu^2}{4(1-\rho_i)}}\right ].
\label{eq: gs_mu}
\end{equation}
Using Eqs.~(\ref{eq: gs_energy}) and (\ref{eq: gs_mu}), we will show that the
mixture of the state with no ions and an electron filling $\rho_e/(1-\rho_i)$
with the state with all ions and no electrons has a lower energy than the 
homogeneous state, i.~e.
\begin{equation}
E(\rho_e,\rho_i)>(1-\rho_i)E(\frac{\rho_e}{1-\rho_i},0)+\rho_iE(0,1).
\label{eq: gs_ineq}
\end{equation}
Moreover, from Eq.~(\ref{eq: gs_ineq}) we will deduce that the mixture 
corresponding to the right hand side of Eq.~(\ref{eq: gs_ineq}) has lower
energy than any other mixture between homogeneous states.
In other words,
\begin{equation}
\alpha E(\rho_e^{\prime},\rho_i^{\prime})+(1-\alpha)E(\rho_e^{\prime\prime},
\rho_i^{\prime\prime})>(1-\rho_i)E(\frac{\rho_e}{1-\rho_i},0)+\rho_iE(0,1),
\label{eq: gs_ineq2}
\end{equation}
where $0\le\alpha\le 1$ and $\rho_e=\alpha\rho_e^{\prime}+(1-\alpha)
\rho_e^{\prime\prime}$, $\rho_i=\alpha\rho_i^{\prime}+(1-\alpha)
\rho_i^{\prime\prime}$, $0<\rho_e^{\prime}<1-\rho_i^{\prime}$, and
$0<\rho_e^{\prime\prime}<1-\rho_i^{\prime\prime}$.  To obtain 
Eq.~(\ref{eq: gs_ineq}),
we first notice that $E(0,1)=0$ and that the chemical potential
$\bar\mu$ corresponding to an electron filling of $\rho_e/(1-\rho_i)$
and an ion filling of zero
is $\bar\mu=\mu/(2\sqrt{1-\rho_i})$, as can be seen from 
Eq.~(\ref{eq: gs_mu}).  Therefore, Eq.~(\ref{eq: gs_energy}) yields
\begin{equation}
(1-\rho_i)E(\frac{\rho_e}{1-\rho_i},0)+\rho_iE(0,1)=
-\frac{4}{3\pi}(1-\rho_i)\left [ 1-
\frac{\mu^2}{4(1-\rho_i)}\right ]^{3/2}
=\frac{1}{\sqrt{1-\rho_i}}E(\rho_e,\rho_i)<E(\rho_e,\rho_i),
\label{eq: gs_maxwell}
\end{equation}
which proves Eq.~(\ref{eq: gs_ineq}).
The proof of Eq.~(\ref{eq: gs_ineq2}) relies on an application of
Eq.~(\ref{eq: gs_ineq})
\begin{eqnarray}
&\alpha& E(\rho_e^{\prime},\rho_i^{\prime})+(1-\alpha)E(\rho_e^{\prime\prime},
\rho_i^{\prime\prime})>\cr
&\alpha& \left [ (1-\rho_i^{\prime})E(\frac{\rho_e^{\prime}}{1-\rho_i^{\prime}},
0)+\rho_i^{\prime}E(0,1)\right ]
+(1-\alpha)
\left [ (1-\rho_i^{\prime\prime})E(\frac{\rho_e^{\prime\prime}}{1-
\rho_i^{\prime\prime}}, 0)+\rho_i^{\prime\prime}E(0,1)\right ].
\label{eq: gs_ineq3}
\end{eqnarray}
The right hand side of Eq.~(\ref{eq: gs_ineq3}) is equal to
\begin{equation}
(1-\rho_i)\left [ \frac{\alpha (1-\rho_i^{\prime})}{1-\rho_i}
E(\frac{\rho_e^{\prime}}{1-\rho_i^{\prime}},0)+
 \frac{(1-\alpha)(1-\rho_i^{\prime\prime})}{1-\rho_i}
E(\frac{\rho_e^{\prime\prime}}{1-\rho_i^{\prime\prime}}, 0)\right ]
+\rho_iE(0,1).
\label{eq: gs_ineq4}
\end{equation}
On the other hand, $E(\rho_e,0)$ is a convex function of $\rho_e$, so
the term inside the brackets in Eq.~(\ref{eq: gs_ineq4}) is greater than
$E(\rho_e/[1-\rho_i],0)$, which yields Eq.~(\ref{eq: gs_ineq}).  We remark
that the convexity of $E(\rho_e,0)$ is obvious from the fact that the free
electron system cannot phase separate.  Formally, it can be seen as follows: 
differentiating Eqs.~(\ref{eq: gs_energy}) and (\ref{eq: gs_mu}) with respect
to $\rho_e$ gives $E^{\prime}(\rho_e,0)=\mu\rho_B(\mu)\partial\mu/\partial
\rho_e$ and $1=\rho_B(\mu)\partial\mu/\partial\rho_e$.  Thus
$E^{\prime}(\rho_e,0)=\mu$ and $E^{\prime\prime}(\rho_e,0)=\partial\mu/
\partial\rho_e=1/\rho_B(\mu)>0$. 

Our interest now is to determine the finite-temperature
phase diagram of the infinite-$U$
Falicov-Kimball model since we know the system always phase separates
at low temperature (although we have not yet ruled out the possibility
of charge-density-wave phases being lower in energy than the phase-separated
ground state). The first step is to evaluate the conduction
electron charge-density-wave susceptibility.  It is often stated that the
Bethe lattice can only support antiferromagnetic or uniform order---no
incommensurate or other ``periodic'' phases can exist.  But this statement
has never been proven, and recent work has shown it to be 
false\cite{bethe_periodic} by a counterexample of a period-three phase
stabilized on the infinite-dimensional Bethe lattice at zero temperature.
The momentum dependence
enters the dressed susceptibility only through the momentum dependence of
the bare susceptibility because the vertex function is local in the
infinite-dimensional limit.  This allows us to simply take
the $U\rightarrow\infty$
of the Brandt-Mielsch result\cite{brandt_mielsch}, which gives
\begin{equation}
1=\rho_i\sum_{n=-\infty}^{\infty}\frac{G_n^2+\chi_n^0(X)}{G_n^2+\rho_i
\chi_n^0(X)},
\label{eq: chitc}
\end{equation}
with $X$ being the parameter that determines the modulation
of the charge-density-wave over the Bethe lattice
and with $\chi_n^0(X)$ the corresponding bare susceptibility.
We do not provide the general formula for all possible charge-density waves
here.  Rather, we present the
three simplifying cases for the susceptibility on the Bethe lattice:
(i) the local susceptibility, where $\chi_n^0(local)=-G_n^2$; (ii)
the ($X=-1$) ``antiferromagnetic'' susceptibility, where
\begin{equation}
\chi_n^0(-1)=-\frac{G_n}{i\omega_n+\mu-\Sigma_n};
\label{eq: chi0bethe-1}
\end{equation}
and (iii) the ($X=1$) uniform susceptibility, where 
\begin{equation}
\chi_n^0(1)=\frac{\partial G_n}{\partial\mu}=-\frac{G_n}{\sqrt{
(i\omega_n+\mu-\Sigma_n)^2-4}}.
\label{eq: chi0bethe1}
\end{equation}
The local susceptibility never has a transition, because the numerator 
of Eq.~(\ref{eq: chitc}) vanishes.  The condition for an ``antiferromagnetic'' 
charge density wave becomes
\begin{equation}
1=\rho_i\sum_{-\infty}^{\infty}\frac{G_n}{i\omega_n+\mu},
\label{eq: tcaf}
\end{equation}
after substituting in the infinite-$U$ form for the self-energy, and using the
quadratic equation $G_n^2-(i\omega_n+\mu)G_n+1-\rho_i=0$ that the interacting
Green's function satisfies.  Now, substituting the integral form for $G_n$
\begin{equation}
G_n=(1-\rho_i)\int_{-\infty}^{\infty}d\epsilon\frac{\rho_B(\epsilon)}
{i\omega_n+\mu-\sqrt{1-\rho_i}\epsilon},
\label{eq: greenintegral}
\end{equation}
into Eq.~(\ref{eq: tcaf}) and performing the sum over Matsubara frequencies
yields the final integral form for $T_c$
\begin{equation}
1=-\frac{\rho_i\sqrt{1-\rho_i}}{2T}\int_{-2\sqrt{1-\rho_i}}^{2\sqrt{1-\rho_i}}
\frac{dz}{z}\frac{\rho_B\left ( \frac{z}{\sqrt{1-\rho_i}}\right )
\tanh\frac{\beta z}{2}}
{\cosh^2\frac{\beta\mu}{2}(1-\tanh\frac{\beta\mu}{2}\tanh\frac{\beta z}{2})},
\label{eq: tcafint}
\end{equation}
(see Appendix A).
But this integrand is positive for all $z$, so the right hand side is always
less than zero, and there is no ``antiferromagnetic'' $T_c$.  The staggered
charge-density-wave order has been found near half-filling 
$\rho_i=\rho_e=1/2$ when the lowest-order exchange for finite-$U$ is
included,\cite{letfulov} but can only occur at $T=0$ and $\rho_i=\rho_e=1/2$
when $U=\infty$.

The uniform susceptibility case is analyzed as follows: First the uniform
susceptibility from Eq.~(\ref{eq: chi0bethe1}) is substituted into 
Eq.~(\ref{eq: chitc}), and the square root is eliminated by using the
exact form for the interacting Green's function in 
Eq.~(\ref{eq: greenbethe}).  Next, the self energy is replaced by its exact form
from Eq.~(\ref{eq: sigmainfinity}), and the quadratic equation for $G_n$ is used
to simplify the $T_c$ equation to
\begin{equation}
1=\rho_i\sum_{n=-\infty}^{\infty}\left [ 1+\frac{1-\rho_i}{(i\omega_n+\mu)
G_n-2(1-\rho_i)}\right ].
\label{eq: tc2}
\end{equation}
Now the interacting form for $G_n$ from Eq.~(\ref{eq: greeninteracting}) is
substituted into Eq.~(\ref{eq: tc2}) and the results simplified to yield
\begin{equation}
1=\rho_i\sum_{n=-\infty}^{\infty}\frac{(i\omega_n+\mu)G_n-2(1-\rho_i)}
{(i\omega_n+\mu)^2-4(1-\rho_i)}.
\label{eq: tc3}
\end{equation}
The final step is to substitute in the integral form for $G_n$ from
Eq.~(\ref{eq: greenintegral}) and perform the summation over Matsubara
frequencies (see Appendix A).  After making a trigonometric substitution, 
the transcendental equation for $T_c$ becomes
\begin{equation}
1=\frac{\rho_i\sqrt{1-\rho_i}}{2\pi T}\int_0^{\pi}d\theta\cos\theta\tanh
\frac{\beta}{2}(2\sqrt{1-\rho_i}\cos\theta-\mu).
\label{eq: tctrans}
\end{equation}

We do not discuss any of the other periodic cases here, because the numerics
involved is cumbersome.  But, we expect the Bethe lattice to have similar 
behavior as the hypercubic lattice, where the transition always went into
the uniform charge-density wave, signifying a phase separation transition.
Details of the other periodic phases will be reported in a future 
publication.

The results for the transition temperature for the uniform charge-density wave
are presented in 
Figure~\ref{fig: bethe_tc}(a).  We choose nine different ion concentrations
ranging from 0.1 to 0.9 in steps of 0.1.  The electron density then
varies from 0 to $1-\rho_i$ for each case.  As can be seen in the figure, the
maximal transition temperature is about $0.12t^*$ and it occurs at 
half-filling of the lower band $\rho_e=(1-\rho_i)/2$ with $\rho_i\approx 0.65$
(coincidentally, this maximal transition temperature is nearly identical to the
maximal $T_c$ to charge-density-wave order at $\rho_e=\rho_i=1/2$ when 
evaluated as a function of the interaction strength $U$).
Since $T_c\ll 1$, we expand Eq.~(\ref{eq: tctrans}) for small $T$ by replacing
the $\tanh x$ by ${\rm sgn} x$ to find
\begin{equation}
T_c\approx\frac{\rho_i\sqrt{1-\rho_i}}{\pi}\sqrt{1-\frac{\mu^2}{4(1-\rho_i)}}.
\label{eq: tcapprox}
\end{equation}
Since the chemical potential will scale with $\sqrt{1-\rho_i}$ for the same
relative electron filling in the lower band $[\rho_e/(1-\rho_i)]$, as shown
in eq.~(\ref{eq: gs_mu}), this form motivates a scaling
plot of $T_c/(\rho_i\sqrt{1-\rho_i})$ versus $\rho_e/(1-\rho_i)$, which
appears in Figure~\ref{fig: bethe_tc}(b).  As can be seen there, the data nearly
collapse on top of each other for $T_c$ {\it which is usually a nonuniversal 
quantity}.  In fact, the variation in $T_c$ is less than 10\% for all different
cases.

The susceptibility analysis shows that the system orders in a uniform
charge-density-wave, which indicates that the system will phase separate
(or segregate) into two regions, one with a higher concentration of electrons
and one with a lower concentration (as we already showed at $T=0$).  
Such a phase separation is usually
associated with a first-order phase transition, rather than a second-order
transition.  Hence, it is important to perform a Maxwell construction of
the free energy that includes mixtures of two states with different electron
and ion concentrations such that $\rho_e=\alpha\rho_e^{\prime}+(1-\alpha)
\rho_e^{\prime\prime}$,
$\rho_i=\alpha\rho_i^{\prime}+(1-\alpha)\rho_i^{\prime\prime}$, and that the 
free energy of the
mixture $F(\rho_e^{\prime},\rho_i^{\prime};\rho_e^{\prime\prime},
\rho_i^{\prime\prime})=\alpha 
F(\rho_e^{\prime},\rho_i^{\prime})+(1-\alpha)F(\rho_e^{\prime\prime},
\rho_i^{\prime\prime})$ is lower in energy than
the pure-phase free energy $F(\rho_e,\rho_i)$.  The second-order phase
transition is the spinodal-decomposition temperature, below which the 
free energy becomes locally unstable in the region of $(\rho_e,\rho_i)$;
in most cases the global free energy is minimized by the Maxwell construction
at a temperature above this spinodal-decomposition temperature.  The
spinodal-decomposition temperature marks the lowest temperature that the system
can be supercooled to before it must undergo a phase transition.

We can calculate the free energy $F(\rho_e,\rho_i)$ for a homogeneous phase
with electron filling $\rho_e$ and ion concentration $\rho_i$ in two 
equivalent ways.  The first method is from Brandt and 
Mielsch\cite{brandt_mielsch} which expresses the free energy in terms of
a summation over Matsubara frequencies as follows:
\begin{eqnarray}
F(\rho_e,\rho_i)&=&-T\ln\frac{1+e^{\beta\mu}}{1-\rho_i}+\int_{-\infty}^{\infty}
d\epsilon\rho(\epsilon)T\sum_{n=-\infty}^{\infty}\ln\left [
\frac{i\omega_n+\mu}{(1-\rho_i)(i\omega_n+\mu-\Sigma_n-\epsilon)}\right ]\cr
&+&\mu\rho_e-\left (T\ln\frac{\rho_i}{1+\rho_i}+T\ln(1+e^{\beta\mu})+T
\sum_{n=-\infty}^{\infty}\ln\left [\frac{1-\rho_i}{(i\omega_n+\mu)G_n}\right ]
\right )\rho_i.
\label{eq: free_bm}
\end{eqnarray}
Similarly, we can evaluate the free energy in the same fashion as Falicov
and Kimball\cite{falicov_kimball} did
\begin{equation}
F(\rho_e,\rho_i)=T\int_{-\infty}^{\infty}d\epsilon
\rho^{int}(\epsilon)\ln\left [ \frac{1}{1+e^{-\beta(\epsilon-\mu)}}\right ]
+T\left [\rho_i\ln\rho_i+(1-\rho_i)\ln(1-\rho_i)\right ]+\mu\rho_e,
\label{eq: free_fk}
\end{equation}
where $\rho^{int}(\epsilon)=\sqrt{4(1-\rho_i)-\epsilon^2}/(2\pi)$ is
the interacting density of states for the Bethe lattice.  We find that both
forms (\ref{eq: free_bm}) and (\ref{eq: free_fk}) are numerically equal
to each other, but are unable to show this result analytically.  Since the
interacting density of states is known for the Bethe lattice, we use the
computationally simpler form in Eq.~(\ref{eq: free_fk}) in our calculations.  
For the hypercubic lattice evaluated in Section III, we employ 
Eq.~(\ref{eq: free_bm}) in the free-energy analysis.

The numerical minimization proceeds in four phases: (i) First a coarse grid is
established for $\rho_i^{\prime}$ and $\rho_i^{\prime\prime}$ and the free 
energy is minimized
over this grid [the electron fillings are determined by the constraints
that the chemical potential is the same in region 1 and region 2 and that 
$\rho_e=\alpha\rho_e^{\prime}+(1-\alpha)\rho_e^{\prime\prime}$, with 
$\alpha$ already determined
from $\rho_i=\alpha\rho_i^{\prime}+(1-\alpha)\rho_i^{\prime\prime}$]; 
(ii) The filling $\rho_i^{\prime\prime}$
is fixed at it's coarse-grid minimal value, and $\rho_i^{\prime}$ is varied on 
a finer
grid to determine the new minimum; (iii) $\rho_i^{\prime}$ is fixed at the new
minimum and $\rho_i^{\prime\prime}$ is now varied on a fine grid to yield a 
new minimal
$\rho_i^{\prime\prime}$; (iv) $\rho_i^{\prime}$ and $\rho_i^{\prime\prime}$ 
are varied together on the same fine
grid to determine the final minimization of the Maxwell construction.  We
found that the minimal values of $\rho_i^{\prime}$ and $\rho_i^{\prime\prime}$ 
rarely changed
in step (iv) confirming the convergence of this method.

We plot our results in Fig.~\ref{fig: bethe_free}.  The first case considered
in Fig.~\ref{fig: bethe_free}(a)
is the case of relative half filling $\rho_e=(1-\rho_i)/2$.  In this case the
chemical potential is always at zero, and the relative electron filling
remains unchanged for all $\rho_i$.  The solid line is the first-order
transition line and the dotted line is the spinodal-decomposition
temperature.  The horizontal distance between the solid lines at a fixed
temperature is a measure of the order parameter $\rho_i^{\prime}-\rho_i^{\prime
\prime}$.  Notice
how the first-order transition temperature is always close to the 
spinodal-decomposition temperature, but that the difference becomes largest at
concentrations close to zero and one.  Note further how the two curves meet
at the maximum (as they must) where the first-order transition disappears
and becomes a second-order transition at a classical critical point.
In Fig.~\ref{fig: bethe_free}(b) we plot the phase diagram for the case
with $\rho_i=0.65$ and $\rho_e=(1-\rho_i)/4$.  In this case the chemical
potential changes as a function of temperature, and as $T\rightarrow 0$
and the system is phase separating into regions with ion densities close
to zero and one, we find that the chemical potential will lie outside of the
bandwidth of the interacting density of states as $\rho_i^{\prime\prime}
\rightarrow 1$ because the bandwidth $(4\sqrt{1-\rho_i})$
becomes narrowed to zero.  In that case, the electron density approaches
zero exponentially fast, which is why the spinodal-decomposition temperature
approaches zero so rapidly in that regime.  For this reason, we find that
the relative electron filling is not a constant in this phase diagram,
as it approaches zero exponentially fast near $\rho_i^{\prime\prime}=1$ and it 
is somewhat larger than quarter filled near $\rho_i^{\prime}=0$.  The two
phase diagrams in Fig.~\ref{fig: bethe_free}(a) and (b)
look similar in the low-density regime, however.  This may
imply that there is an analogous scaling regime for the first-order $T_c$,
but it does not look like there would be a universal curve for the region
close to $\rho_i^{\prime\prime}=1$.  The numerical effort required to perform 
the free-energy analysis is significant, so a thorough analysis of any possible
scaling forms for $T_c$ was not performed.

\section{Results for the hypercubic lattice}

The formalism for the hypercubic lattice is essentially unchanged from
the Bethe lattice.  The main differences are that the integrals cannot
be performed analytically anymore, which requires results to be worked out
numerically, and requires more computational effort.  The basic framework
in Eqs.~(\ref{eq: greendef})---(\ref{eq: sigmainfinity}) is identical
as before, except now the noninteracting density of states is a Gaussian
for the hypercubic lattice.  The integral for the local Green's function
is no longer elementary, and so one needs to solve the problem iteratively
as was done previously for the Falicov-Kimball model: (i) first the 
self energy is set equal to zero; (ii) next the local Green's function is 
determined from Eq.~(\ref{eq: greenlocal}); (iii) then the self energy is
determined from Eq.~(\ref{eq: sigmainfinity}); (iv) then steps (ii) and (iii)
are repeated until the equations converge.  We can compare the results
of the interacting Green's function to the form found before for the Bethe
lattice, by approximating the interacting density of states in the same
fashion as before:  we narrow the Gaussian by the factor $\sqrt{1-\rho_i}$
and have a total weight of $1-\rho_i$ in the density of states.  When we
compare the Green's function along the imaginary axis at half filling
in Fig.~\ref{fig: hyp_green}(a) we find that that approximation works well
at high energies, but begins to fail near zero frequency (we chose $\rho_e=1/6$,
$\rho_i=2/3$, and $T=0.1$). The solid line is the exact result and the
dotted line is the approximate (band-narrowed) result.  The infinite-$U$
Green's function on the hypercubic lattice is more complicated than on the Bethe
lattice and the simple form that describes it for the Bethe lattice
no longer holds.  This is the main reason why the hypercubic lattice
is more complicated to deal with than the Bethe lattice. To see this more
fully, we examine the interacting density of states in 
Fig.~\ref{fig: hyp_green}(b). The interacting density of states is determined
by solving the same equations for the Green's function, but this time on the
real axis, rather than the imaginary axis. Notice how the band-narrowed form 
$\sqrt{1-\rho_i}\exp(-\epsilon^2/[1-\rho_i])/\sqrt{\pi}$ (dotted line)
is a reasonable approximation to the interacting density of states (solid line) 
but that it is too narrow, and it overestimates the peak height.

Since we do not know an analytic form for
the interacting density of states, we cannot 
perform the same kind of analysis that we did before at zero temperature to 
see if the system is phase separated.  But we can examine the finite-temperature
phase diagrams in the same manner.  The susceptibility diverges whenever
Eq.~(\ref{eq: chitc}) is satisfied.  The bare susceptibility now
takes the form
\begin{equation}
\chi_n^0(X)=-\frac{1}{2\pi}\int_{-\infty}^{\infty}\rho(y)\frac{1}{i\omega_n
+\mu-\Sigma_n-y}\int_{-\infty}^{\infty}dz\rho(z)\frac{1}{i\omega_n+\mu-\Sigma_n
-yX-z\sqrt{1-X^2}},
\label{eq: chi0_hyp}
\end{equation}
where $X=\lim_{d\rightarrow\infty}\sum_{j=1}^d\cos(k_j)$ for the ordering
wavevector {\bf k}.  The bare susceptibility continues to assume a simple
form for the same three cases: (i) $X=0$ the local susceptibility where
$\chi_n^0(0)=-G_n^2$; (ii) $X=-1$ the ``antiferromagnetic'' susceptibility,
where $\chi_n^0(-1)=-G_n/(i\omega_n+\mu-\Sigma_n)$; and (iii) $X=1$ the
uniform susceptibility, where
\begin{equation}
\chi_n^0(1)=\frac{\partial G_n}{\partial\mu}=2[1-(i\omega_n+\mu-\Sigma_n)G_n].
\label{eq: chi0_uniform_hyp}
\end{equation}

If we try to approximate the transition temperature by substituting in the
approximate form we have for $G_n$ derived by assuming the interacting
density of states has the same shape, but is band narrowed, we find that
the $T_c$'s generated are not accurate at all.  Hence, the simple 
band-narrowing approximation
works reasonably well for the Green's function, but is poor for the
susceptibility.

Instead, we simply solve for the transition temperatures numerically.  We
find in every case that we examined that the transition temperature is
always highest for $X=1$ and vanishes for all $X\le 0$.  This is shown in
Fig.~\ref{fig: tc_x} for the cases of $\rho_e=1/6$, 1/12, 1/24, 1/48, 1/96,
1/192, 1/384, and 1/768 and $\rho_i=2/3$, which ranges from relative
half filling to the low-density regime.  We
plot $T_c(X)$ and see that the system always favors the uniform charge-density
wave, signifying that the system wants to phase separate.  We calculate the
spinodal-decomposition temperature for phase separation by finding the
temperature at which the uniform susceptibility diverges as a function
of $\rho_e$ and $\rho_i$.  These temperatures are plotted in 
Fig.~\ref{fig: tc_hyp_spinodal}(a).  This plot looks similar to
what we found for the Bethe lattice before, so we try the same scaling
form in Fig.~\ref{fig: tc_hyp_spinodal}(b), 
plotting $T_c/(\rho_i\sqrt{1-\rho_i})$ versus $\rho_e/(1-\rho_i)$.
Once again we see a data collapse, but the spread in the $T_c$'s is somewhat
larger than that  seen in the Bethe lattice.

Finally, we calculate the full phase diagram for the case of relative
half filling $\rho_e=(1-\rho_i)/2$ where $\mu=0$ in Fig.~\ref{fig: tc_hyp_free}.
The form of this result is similar to what was seen in the Bethe lattice.
The first-order transition temperature and the spinodal-decomposition 
temperature meet at the peak of the curve where the first-order transition
becomes second order.  We did not perform a free-energy calculation at
other relative
fillings here, because the numerical solution was significantly more
difficult due to the fact that we needed to use Eq.~(\ref{eq: free_bm}) rather
than the computationally simpler  Eq.~(\ref{eq: free_fk}).

\section{Conclusions and Discussion}

We have provided an exact solution to the spinless Falicov-Kimball model in
the strongly correlated limit of $U=\infty$.  We only considered the 
less than unit density cases $0<\rho_e<1-\rho_i$, because they all satisfy
a similar functional form.  On the Bethe lattice we found that the 
system always phase separated at $T=0$ to states where the electrons all moved
to one part of the lattice, and the ions moved to the other part.  The
spinodal-decomposition temperature for segregation solved a simple 
transcendental equation, which we showed collapsed onto a scaling curve. 
In addition, we solved for the first-order transition temperature for 
a select number of cases and discovered that the first-order transition usually
occurred quite close to the spinodal-decomposition temperature.  On the
hypercubic lattice, we found similar results, but had to carry the analysis out
numerically for all cases considered.  We were able to explicitly show that
phase separation precluded incommensurate (or commensurate) charge-density-wave
order for the hypercubic lattice.

These results show that when the screened Coulomb interaction is large,
or in the alloy picture, when the A ions are extremely different from the
B ions, then the system will segregate at low temperatures.  This proves
the segregation principle for the infinite-dimensional limit, and leads
us to believe that it holds for all dimensions (since it has also been
demonstrated in one dimension).  Future problems to be investigated include
a study of the case $\rho_e+\rho_i=1$, as well as finite Coulomb interaction
$U$.  In the unit-density
case, we expect charge-density-wave order to be more prevalent, perhaps
precluding the segregated phase for all $U$.  When the strength of
the Coulomb interaction is reduced, we expect the segregated phase to
gradually disappear and be taken over by other phase-separated or 
charge-density-wave ordered phases.

It is possible that the phenomena described here incorporates the relevant 
physics of the charge-stripe phases in the cuprate materials:  that the
stripes occurred because of the strong propensity towards phase separation
in the strongly correlated limit.  The analogy would stem from considering
the down spin electrons of the Hubbard model to be frozen in a particular
configuration, and then examine how the mobile up spin electrons react
to the down spins.  The quantum fluctuations of the Hubbard model are
replaced by the thermal fluctuations of the Falicov-Kimball model, and
it can be viewed as a simplifying approximation to the charge dynamics of
the strongly correlated Hubbard model, but not incorporating the spin
dynamics.  In this case, as postulated by
Emery and Kivelson,\cite{emery_kivelson} the stripes would form from a
balance between the desire for the system to phase separate, and the 
long-range Coulomb interaction, which would prevent the electrons from
completely separating from the ions.  There is evidence for alternative
points of view, however.  White and collaborators\cite{white} have shown
that the Hubbard model on a ladder displays charge-stripe order even without
the long-range Coulomb interaction.  This order arises from  the correlation
of the spins and the holes and a desire to reduce the frustration induced
by the hole motion.  In their picture, the stripe ordering arises completely
from a model that includes no long-range forces.  Nevertheless,
it is our belief that
the phase separation exhibited here will be an important element of a
complete description of the charge-stripe order in the cuprates and
nickelates, because it must occur if $U$ becomes large enough.

\acknowledgments

J.K.F. acknowledges support of this work from
the Office of Naval Research Young Investigator Program N000149610828.
J.K.F. also acknowledges the hospitality he received at the Institut de
Physique Theorique in Lausanne, where this work was initiated.

\appendix

\section{Derivation of the transcendental equations for $T_c$ on the
Bethe lattice }

The derivation of Eqs.~(\ref{eq: tcafint}) and (\ref{eq: tctrans})
involve summations over Matsubara frequencies, which are performed with the
help of the identity
\begin{equation}
\tanh \frac{\beta x}{2}=\frac{2}{\beta}\sum_{n=-\infty}^{\infty}\frac{1}
{i\omega_n+x},
\label{eq: tanhident}
\end{equation}
for any real number $x$.  Using Eq.~(\ref{eq: greenintegral}) and the change of
variables $\epsilon\rightarrow\epsilon/\sqrt{1-\rho_i}$ yields
\begin{equation}
\frac{G_n}{i\omega_n+\mu}=\sqrt{1-\rho_i}\int_{-\infty}^{\infty}
\frac{d\epsilon}{\epsilon}\rho_B(\frac{\epsilon}{\sqrt{1-\rho_i}})\left [
\frac{1}{i\omega_n+\mu-\epsilon}-\frac{1}{i\omega_n+\mu}\right ].
\label{eq: af1}
\end{equation}
Employing Eq.~(\ref{eq: tanhident}) then shows that
\begin{equation}
\sum_{n=-\infty}^{\infty}\frac{G_n}{i\omega_n+\mu}=
\frac{2\sqrt{1-\rho_i}}{\beta}
\int_{-\infty}^{\infty}\frac{d\epsilon}{\epsilon}\rho_B(\frac{\epsilon}
{\sqrt{1-\rho_i}})\left [ \tanh\frac{\beta(\mu-\epsilon)}{2}-
\tanh\frac{\beta\mu}{2}\right ].
\label{eq: af2}
\end{equation}
Eq.~(\ref{eq: tcafint}) then follows from the trigonometric identity
\begin{equation}
\tanh\frac{\beta(\mu-\epsilon)}{2}-\tanh\frac{\beta\mu}{2}=
\frac{\tanh\frac{\beta\mu}{2}}{\cosh^2\frac{\beta\mu}{2}
(1-\tanh\frac{\beta\mu}{2}\tanh\frac{\beta\epsilon}{2})}.
\label{eq: trigident}
\end{equation}

The derivation of Eq.~(\ref{eq: tctrans}) is more involved, but proceeds 
along the same lines.  Using the integral representation in 
Eq.~(\ref{eq: greenintegral}) for $G_n$, performing a decomposition into simple
fractions, and then using the identity in Eq.~(\ref{eq: tanhident})
produces both
\begin{eqnarray}
\sum_{n=-\infty}^{\infty}\frac{(i\omega_n+\mu)G_n}{(i\omega_n+\mu)^2-4
(1-\rho_i)}&=&
\frac{\beta}{4}\sqrt{1-\rho_i}\tanh\frac{\beta(\mu-2\sqrt{1-\rho_i})}{2}
\int_{-\infty}^{\infty}d\epsilon\frac{\rho_B(\epsilon)}{2-\epsilon}\cr
&-&\frac{\beta}{4}\sqrt{1-\rho_i}\tanh\frac{\beta(\mu+2\sqrt{1-\rho_i})}{2}
\int_{-\infty}^{\infty}d\epsilon\frac{\rho_B(\epsilon)}{2+\epsilon}\cr
&+&\frac{\beta}{4}\sqrt{1-\rho_i}\int_{-\infty}^{\infty}d\epsilon
\left [\frac{\rho_B(\epsilon)}{2+\epsilon}-\frac{\rho_B(\epsilon)}{2-\epsilon}
\right ]
\tanh\frac{\beta(\mu-\epsilon\sqrt{1-\rho_i})}{2},
\label{eq: f1}
\end{eqnarray}
and
\begin{equation}
\sum_{n=-\infty}^{\infty}\frac{2(1-\rho_i)}{(i\omega_n+\mu)^2-4
(1-\rho_i)}=\frac{\beta}{4}\sqrt{1-\rho_i}\left [
\tanh\frac{\beta(\mu-2\sqrt{1-\rho_i})}{2}-
\tanh\frac{\beta(\mu+2\sqrt{1-\rho_i})}{2}\right ].
\label{eq: f2}
\end{equation}
Now we use the fact that the integrals for the noninteracting Green's
function are trivial
\begin{equation}
\int_{-\infty}^{\infty}d\epsilon\frac{\rho_B(\epsilon)}{2-\epsilon}=
\int_{-\infty}^{\infty}d\epsilon\frac{\rho_B(\epsilon)}{2+\epsilon}=1,
\label{eq: intident}
\end{equation}
and subtract Eq.~(\ref{eq: f1}) from Eq.~(\ref{eq: f2}) to obtain
\begin{equation}
\sum_{n=-\infty}^{\infty}\frac{(i\omega_n+\mu)G_n-2(1-\rho_i)}
{(i\omega_n+\mu)^2-4(1-\rho_i)}=\frac{\beta}{4}\sqrt{1-\rho_i}\int_{-2}^{2}
d\epsilon \frac{\epsilon}{\pi\sqrt{4-\epsilon^2}}\tanh
\frac{\beta(\epsilon\sqrt{1-\rho_i}-\mu)}{2}.
\label{eq: final}
\end{equation}
Eq.~(\ref{eq: tctrans}) then follows from the change of variables
$\epsilon=2\cos\theta$.

\begin{figure}[t]
\caption{Second-order transition temperature on the Bethe lattice
(corresponding to spinodal decomposition). (a) Transition temperature
plotted as a function of electron filling. (b) Transition temperature
plotted on a scaling curve as a function of relative electron filling.}
\label{fig: bethe_tc}
\end{figure}

\begin{figure}[t]
\caption{Transition temperature to phase separation on the Bethe lattice.
(a) The case of relative half-filling $(\rho_e=[1-\rho_i]/2)$.  The solid
line is the first-order transition temperature and the dotted line is
the spinodal decomposition temperature.  Notice how these two curves meet
at the maximum where the first-order transition becomes second order.
(b) The case near relative quarter filling (as described in the text).
Notice how the shape of the curve differs from (a) near $\rho_i=1$.  This
is because the electron filling becomes exponentially small once the chemical
potential lies outside of the interacting bandwidth. }
\label{fig: bethe_free}
\end{figure}

\begin{figure}[t]
\caption{Comparison of the band-narrowed approximation with the exact
result on the hypercubic lattice.  (a) The Green's function at relative
half filling on the imaginary axis.  The solid line is the exact result
and the dotted line is the approximation.  The parameters are $\rho_e=1/6$,
$\rho_i=2/3$, and $T=0.1$. (b) The interacting density of states (which
is temperature-independent) for the exact (solid line) and approximate 
(dotted line) cases with $\rho_e=1/6$ and $\rho_i=2/3$}
\label{fig: hyp_green}
\end{figure}

\begin{figure}[t]
\caption{Plot of the transition temperature versus the ordering wave vector
$X(k)$ for the case $\rho_i=2/3$ and values of $\rho_e$ ranging from
relative half filling ($\rho_e=1/6$) top curve, to the low-density regime
($\rho_e=1/784$) bottom curve, with the density reduced by a factor of 2
for each case.}
\label{fig: tc_x}
\end{figure}

\begin{figure}[t]
\caption{Second-order transition temperature on the hypercubic lattice
(corresponding to spinodal decomposition). (a) Transition temperature
plotted as a function of electron filling. (b) Transition temperature
plotted on a scaling curve as a function of relative electron filling.}
\label{fig: tc_hyp_spinodal}
\end{figure}

\begin{figure}[t]
\caption{Transition temperature to phase separation on the hypercubic lattice
for the case of relative half-filling $(\rho_e=[1-\rho_i]/2)$.  The solid
line is the first-order transition temperature and the dotted line is
the spinodal decomposition temperature.  Notice how these two curves meet
at the maximum where the first-order transition becomes second order.}
\label{fig: tc_hyp_free}
\end{figure}

\begin{figure}[tbp]
\centerline{
\epsfxsize=2.5in \epsffile{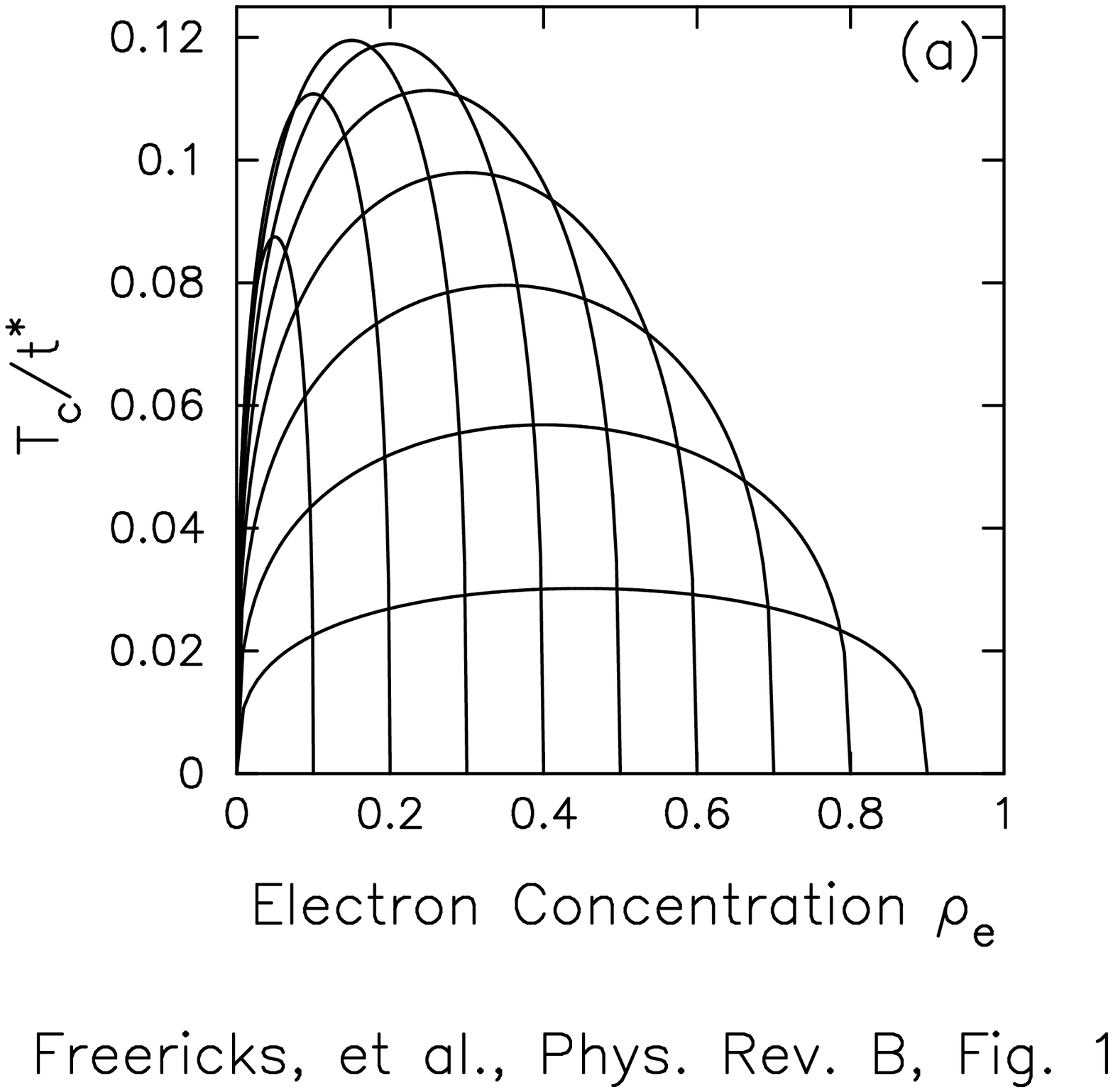} \epsfxsize=2.5in \epsffile{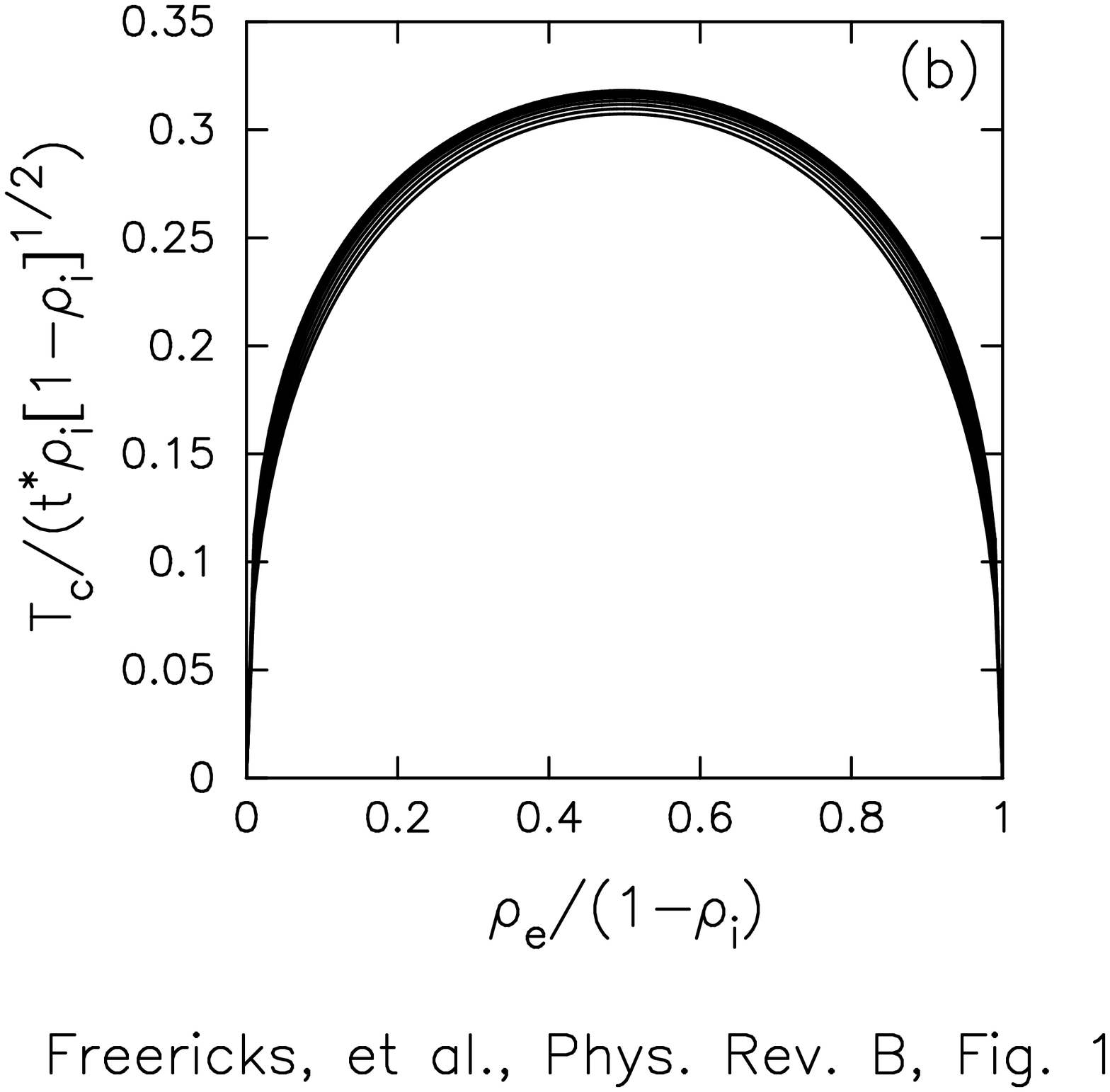}}
\end{figure}

\begin{figure}[tbp]
\centerline{
\epsfxsize=2.5in \epsffile{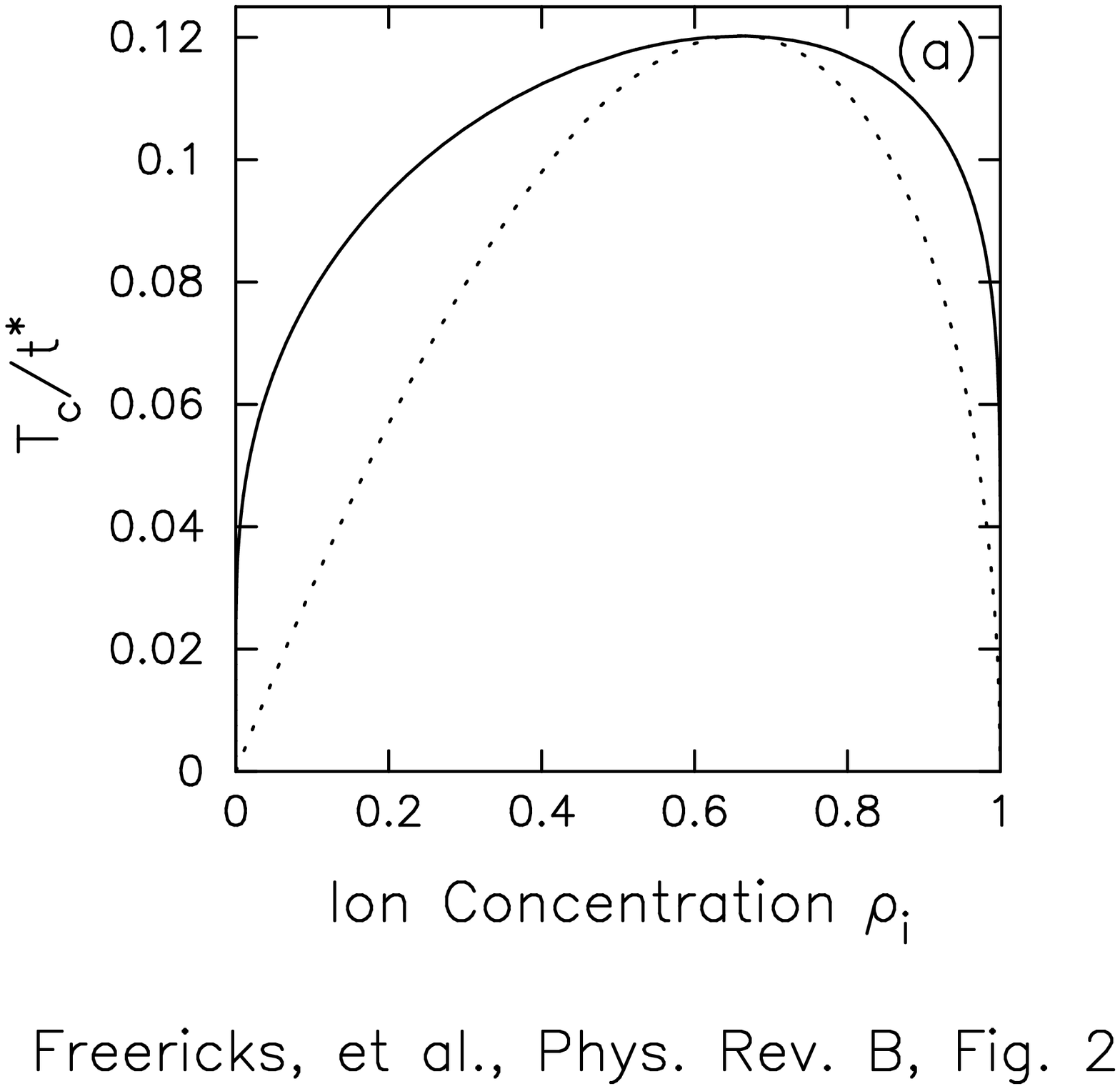} \epsfxsize=2.5in \epsffile{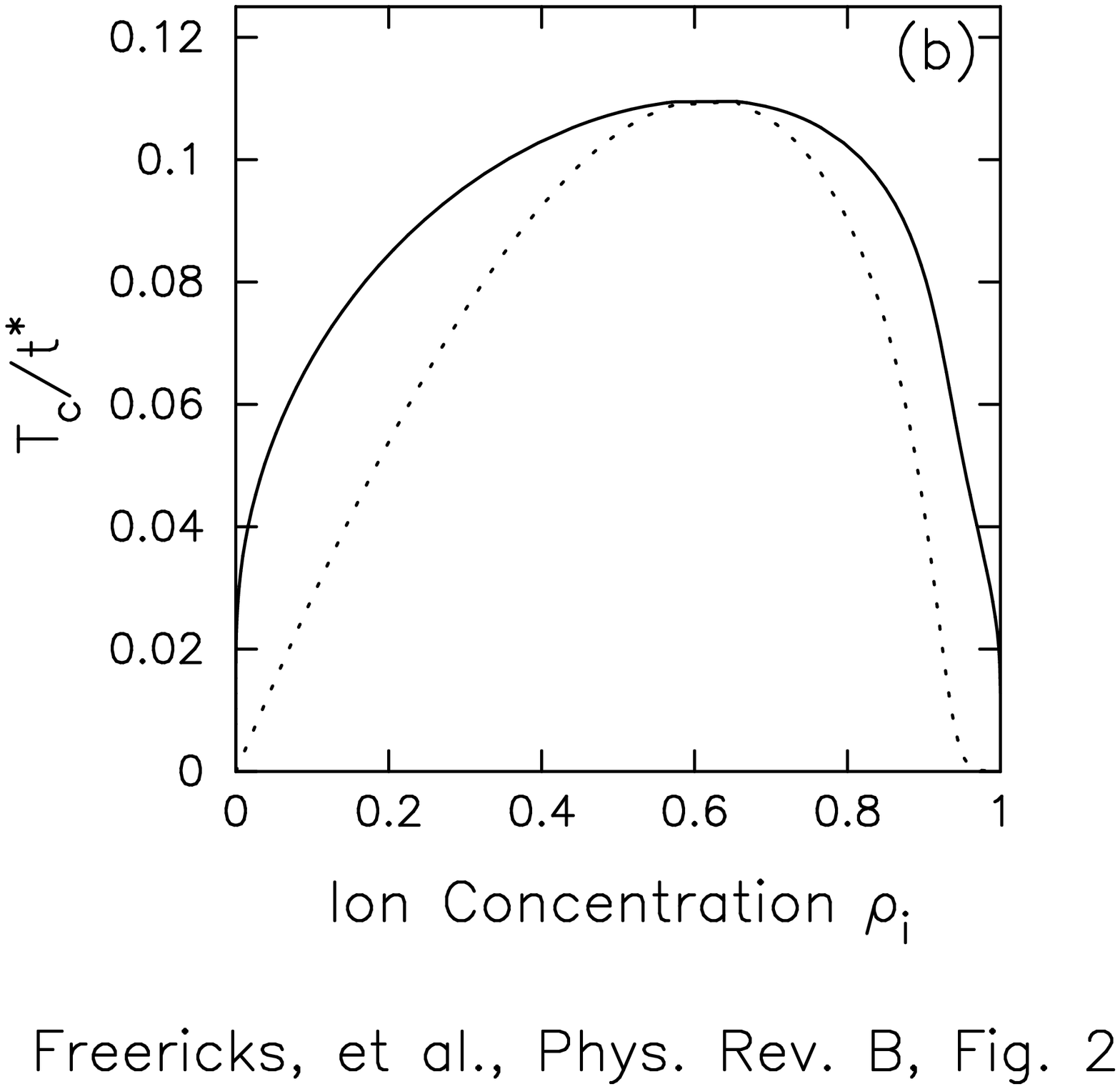}}
\end{figure}

\begin{figure}[tbp]
\centerline{
\epsfxsize=2.5in \epsffile{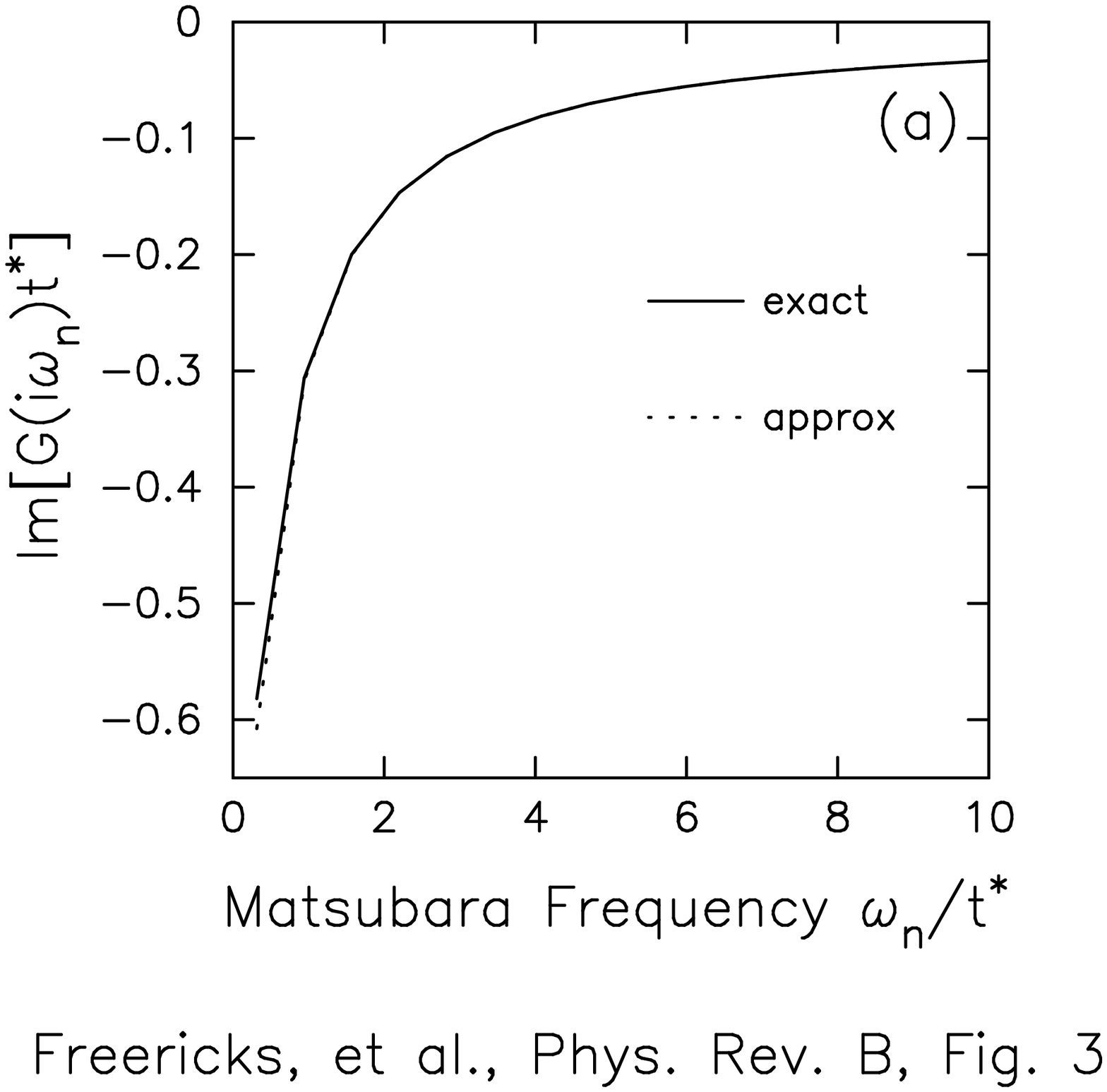} \epsfxsize=2.5in \epsffile{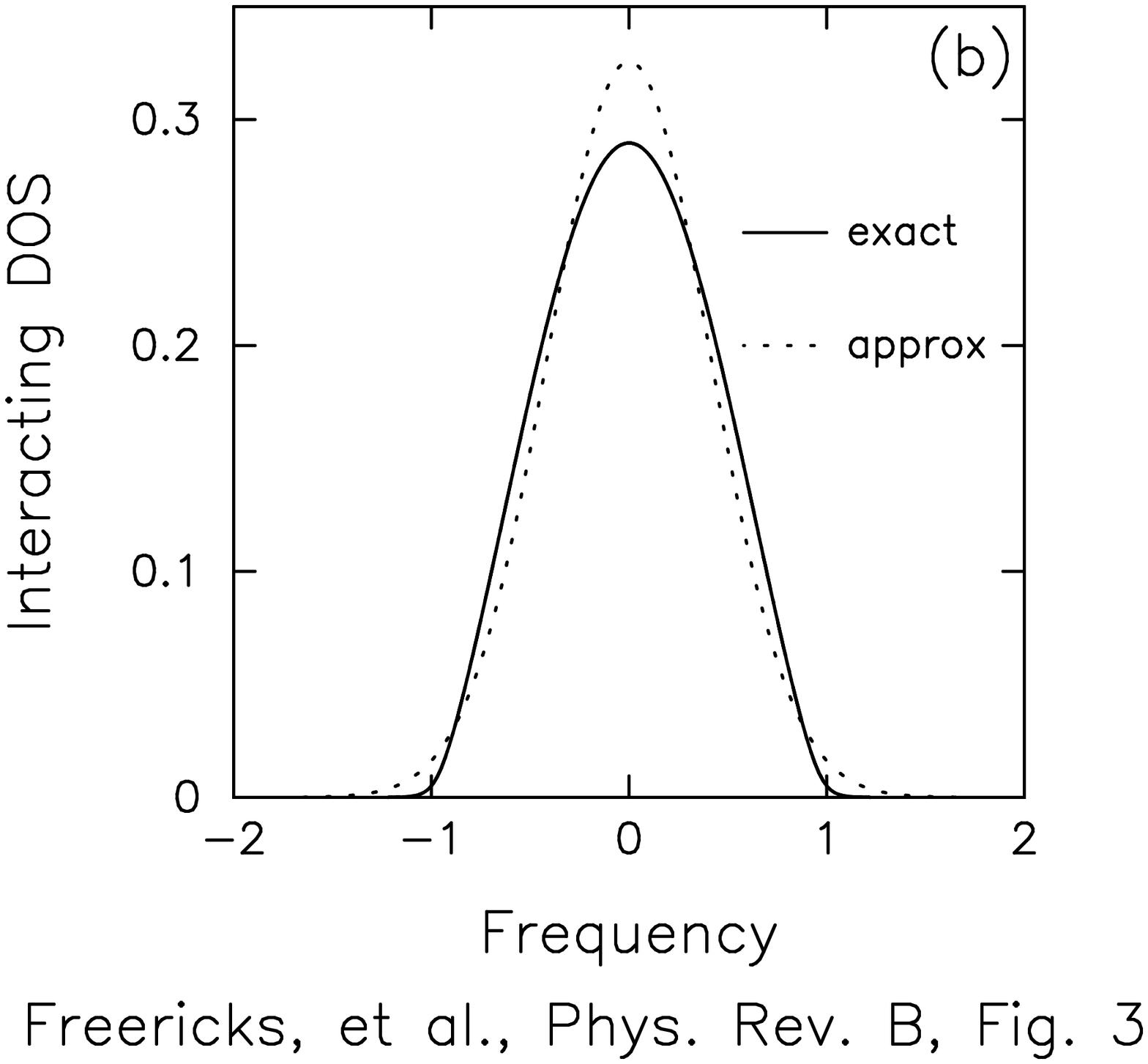} }
\end{figure}

\begin{figure}[tbp]
\centerline{
\epsfxsize=2.5in \epsffile{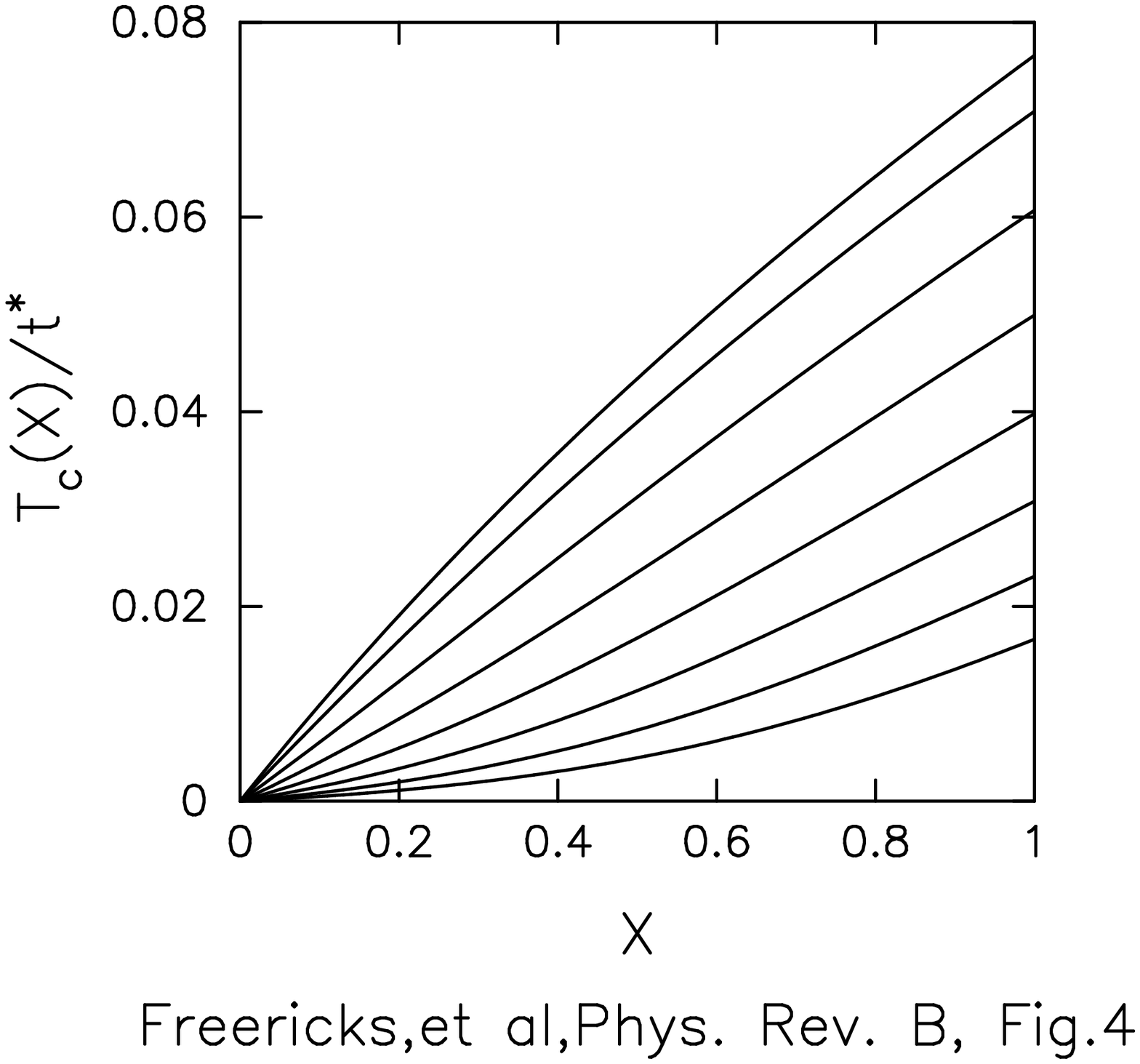}}
\end{figure}

\begin{figure}[tbp]
\centerline{
\epsfxsize=2.5in \epsffile{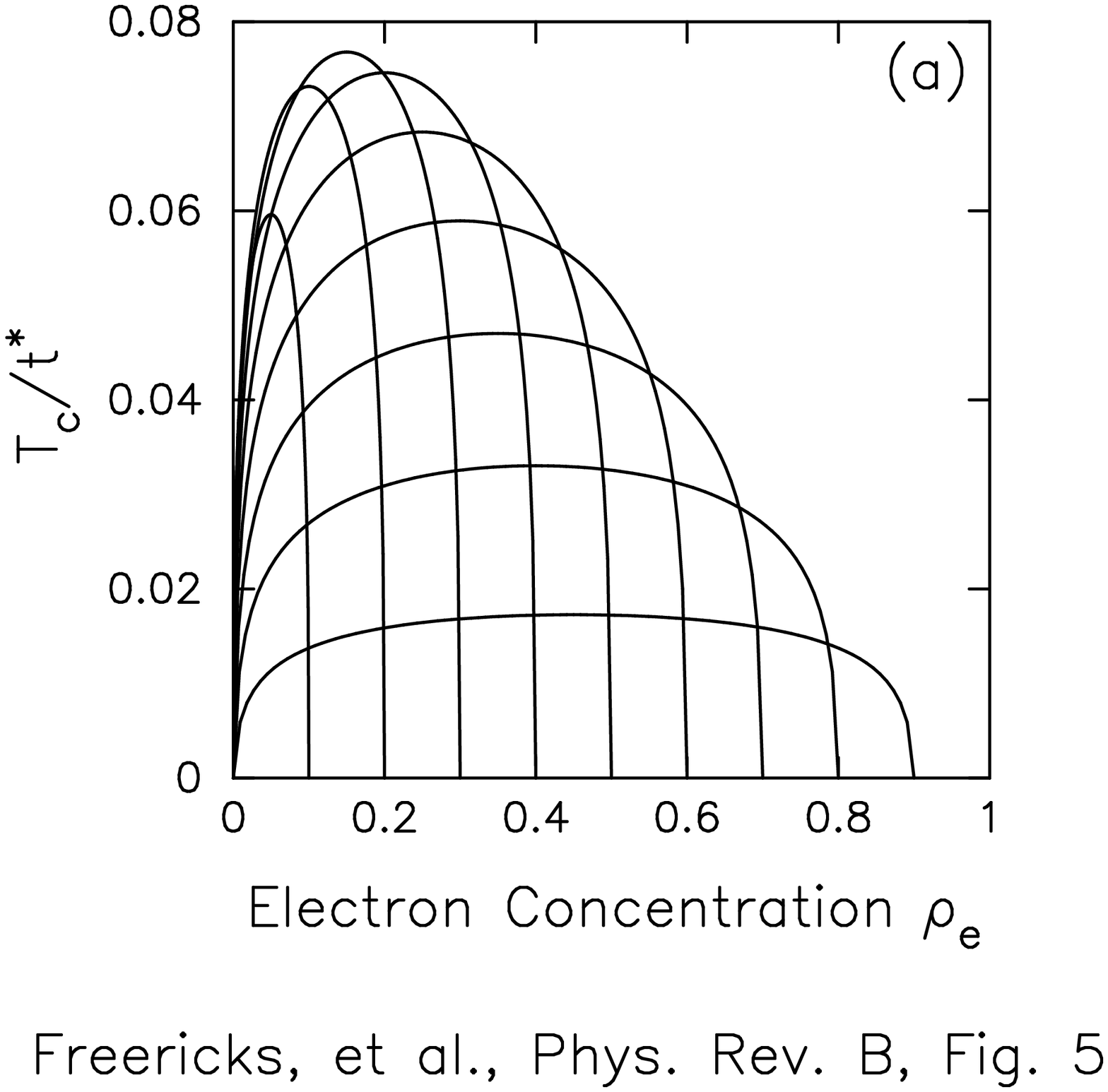} \epsfxsize=2.5in \epsffile{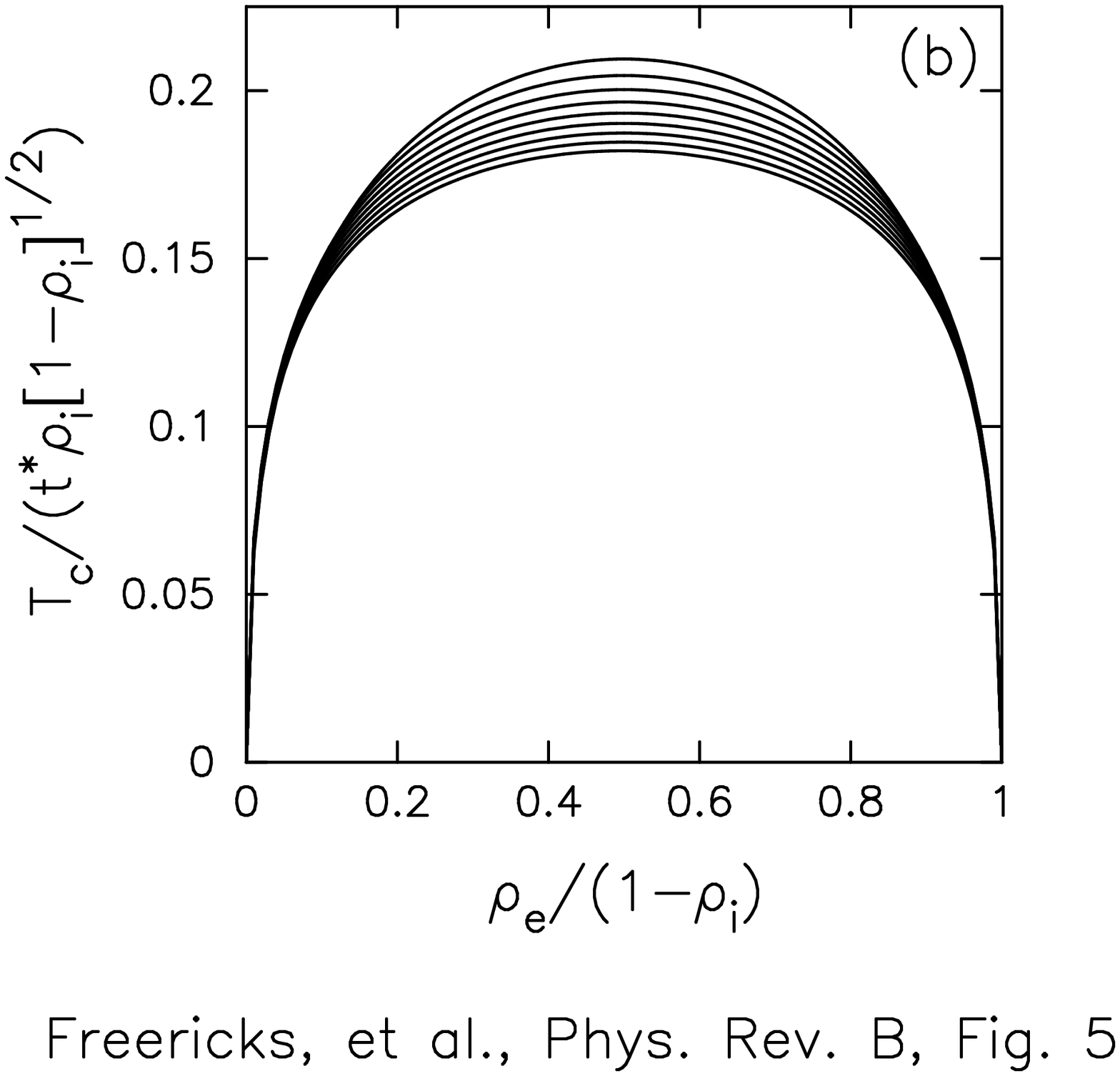}}
\end{figure}

\begin{figure}[tbp]
\centerline{
\epsfxsize=2.5in \epsffile{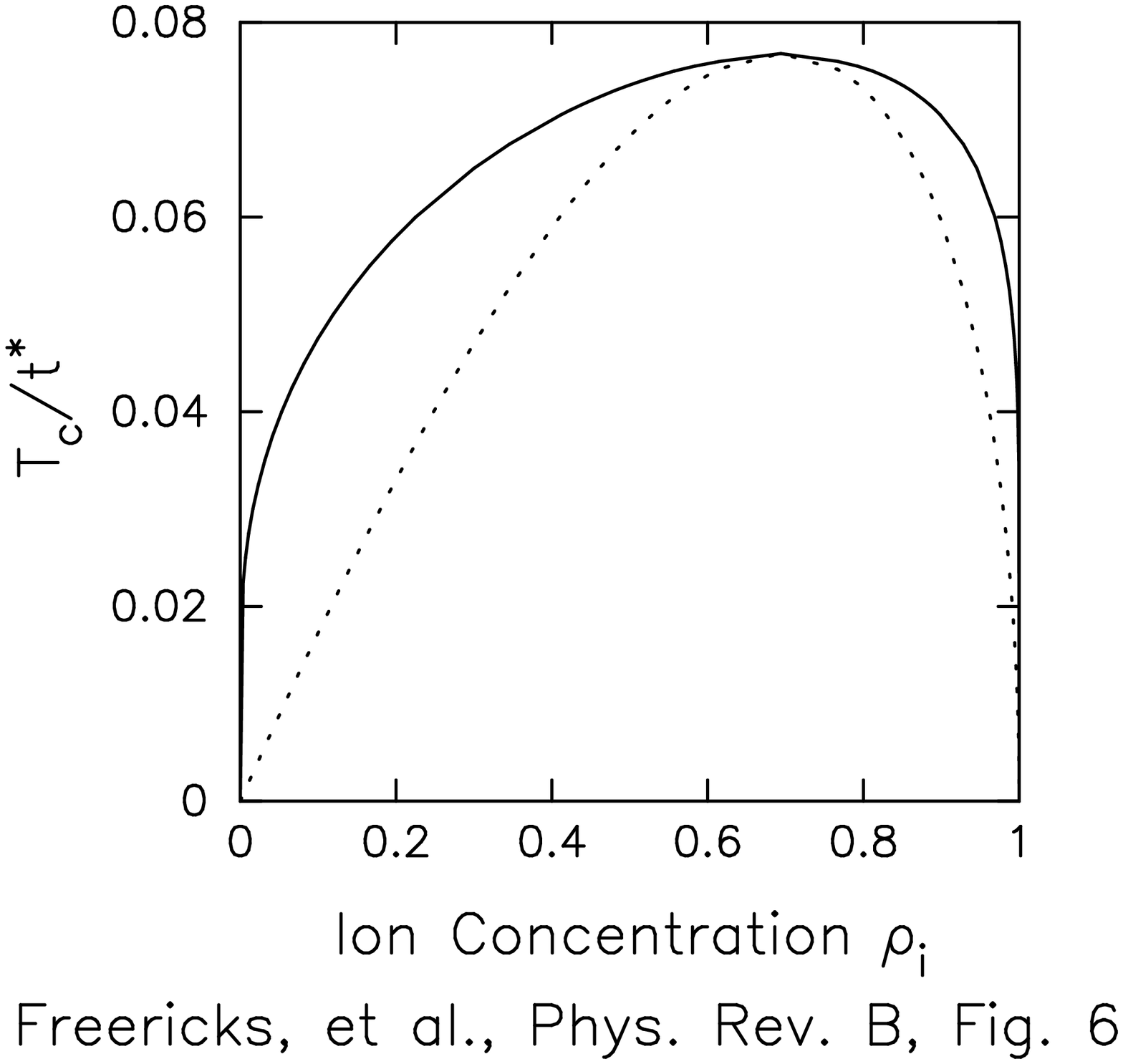}}
\end{figure}

\end{document}